\documentclass[a4paper,11pt]{article}
\usepackage{jinstpub} 
\usepackage{lineno}
\usepackage{subcaption}
\usepackage{multirow}

\title{\boldmath Optical readout of MPGDs with solid wavelength shifters}

\author[a]{F.M.~Brunbauer}
\author[b,1]{A.~Cools\note{Now at The University of Melbourne, School of Physics, The University of Melbourne, Melbourne, Victoria 3010, Australia}}
\author[c]{M.~Cortesi}
\author[b]{E.~Fasoula}
\author[b]{E.~Ferrer-Ribas}
\author[a]{K.J.~Flöthner} 
\author[d]{F.~Garcia} 
\author[a]{D.~Janssens} 
\author[a]{M.~Lisowska}
\author[a]{P.~Sviatopolk Mirsky}
\author[a,e]{H.~Müller} 
\author[a,d,f]{J.~Nummi}
\author[a]{E.~Oliveri} 
\author[a]{G.~Orlandini} 
\author[b]{T.~Papaevangelou} 
\author[a,g]{D.~Pfeiffer} 
\author[b]{E.~Pollacco} 
\author[a]{L.~Ropelewski} 
\author[a]{F.~Sauli}
\author[g]{J.~Samarati} 
\author[a]{L.~Scharenberg} 
\author[a]{M.~van Stenis} 
\author[a]{R.~Veenhof}

\affiliation[a]{European Organization for Nuclear Research (CERN), CH-1211 Geneve 23, Switzerland}
\affiliation[b]{Institut de Recherche sur les lois Fondamentales de l’Univers (IRFU, CEA), Université Paris Saclay, F-91191 Gif-sur-Yvette, France}
\affiliation[c]{Facility for Rare Isotope Beams, 640 South Shaw Lane, East Lansing, MI 48824, USA}
\affiliation[d]{Helsinki Institute of Physics, University of Helsinki, Gustaf Hällströmin katu 2, 00560 Helsinki, Finland}
\affiliation[e]{University of Bonn, Regina-Pacis-Weg 3, 53113 Bonn, Germany}
\affiliation[f]{Aalto University, School of Science, Department of Applied Physics, Otakaari 24, 02150 Espoo, Finland}
\affiliation[g]{European Spallation Source (ESS ERIC), P.O. Box 176, SE-22100 Lund, Sweden}

\emailAdd{florian.brunbauer@cern.ch}

\abstract{
Optical readout of MicroPattern Gaseous Detectors (MPGDs) makes use of the high granularity of imaging sensors to achieve good spatial resolution for radiation imaging and particle detection. CF$_{4}$ is widely used as a scintillating gas because its emission lies in the visible range, where optical sensors are most sensitive. However, to reduce reliance on greenhouse gas, such as CF$_{4}$, which also has limited availability, alternative gas mixtures emitting scintillation light in the ultraviolet range can be used in combination with wavelength shifters.
We investigate the spatial resolution achievable with optically read out Gaseous Electron Multipliers (GEMs) and Micromegas when using solid wavelength shifter layers such as Tetraphenyl butadiene (TPB). TPB coatings on the anode of glass Micromegas achieve the best spatial resolution of 0.22\,mm owing to the minimal distance between the origin of the scintillation light and the wavelength shifter. Nevertheless, TPB layers were also shown to achieve moderate spatial resolution in combination with optically read out GEMs.
}


\begin{document}

\maketitle

\flushbottom

\section{Introduction}
\label{sec:intro}

Recording scintillation light emitted during avalanche amplification in gaseous radiation detectors is a performant alternative  approach compared to the readout of induced electrical signals. The so-called optical readout method can be used to acquire high-granularity images exploiting the high pixel count of state-of-the-art imaging sensors in commercially available cameras.

The main requirement for optical readout is a good match between the scintillation light emission spectrum of gas mixtures and the spectral efficiency of recording devices. While most gases emit in the ultraviolet range, the majority of cameras are optimised for sensitivity in the visible wavelength range. A notable exception is carbon tetrafluoride (CF$_{4}$), which features abundant visible scintillation yield and has become a popular choice for optically read out gaseous detectors either as pure gas or in mixtures with noble gases. Scintillation spectra of mixtures such as Ar/CF$_{4}$ or He/CF$_{4}$ feature molecular scintillation emission bands of CF$_3$ fragments in the UV and visible ranges \cite{cf4GEMScintillation}.

While CF$_{4}$ is an attractive choice for many gaseous detectors including its use in the GEM-based optical TPCs used by the MIGDAL \cite{migdalExperiment} and CYGNO \cite{cygnoExperiment} experiments, it may not fit technical or physics requirements of other applications. In addition, due to its high global warming potential, the use of CF$_{4}$ is deprecated and its availability is decreasing.    

To enable the use of the optical readout approach of gaseous detectors with gas mixtures which do not contain CF$_{4}$, recording devices compatible with the emitted light spectrum must be used or emitted light may be converted to a suitable wavelength range with wavelength shifting materials. Tetraphenyl butadiene (TPB) is a popular solid wavelength shifter for converting EUV and UV light to the visible range.
With absorption for a range of wavelengths from VUV to near UV and a strong re-emission peak centered around 425\,nm \cite{tpbEmissionAbsorptionLiquidArTPCs}, TPB is well suited to convert short wavelength incident light to a visible range compatible with conventional imaging sensors.
A high total fluorescence efficiency has been demonstrated for TPB over a range of wavelengths with a maximum total efficiency of approximately 1.2 at 128\,nm \cite{Gehman_2011}. Over time, TPB ages in ambient conditions and its conversion efficiency decreases which is particularly noticable for vapor-deposited TPB layers \cite{tpbEfficiencyGraybill}. TPB exhibits a high surface resistivity on the order of $10^{13}\, \Omega / \square$.

TPB and PolyEthylene Naphthalate (PEN) are potential options of solid wavelength shifters from UV wavelengths to the visible range. In direct comparison, PEN features a lower conversion efficiency of around 34\% to 47\% of the efficiency of TPB \cite{TPBPENComparison}. While TPB is deposited by vacuum evaporation and thin films are used for wavelength shifting, PEN foils are readily available and may offer advantages where the use of thin films may not be possible.

\section{Experimental methods}
\label{sec:setup}

The spatial resolution of a GEM \cite{SAULI1997531} and Micromegas \cite{GIOMATARIS199629} detectors employing TPB thin films as a wavelength shifters (WLSs) was studied to compare the effect of the location of the WLS on the achievable spatial resolution.

\begin{figure*}[t!]
    \centering
    \begin{subfigure}[t]{0.5\textwidth}
        \centering
        \includegraphics[height=2in]{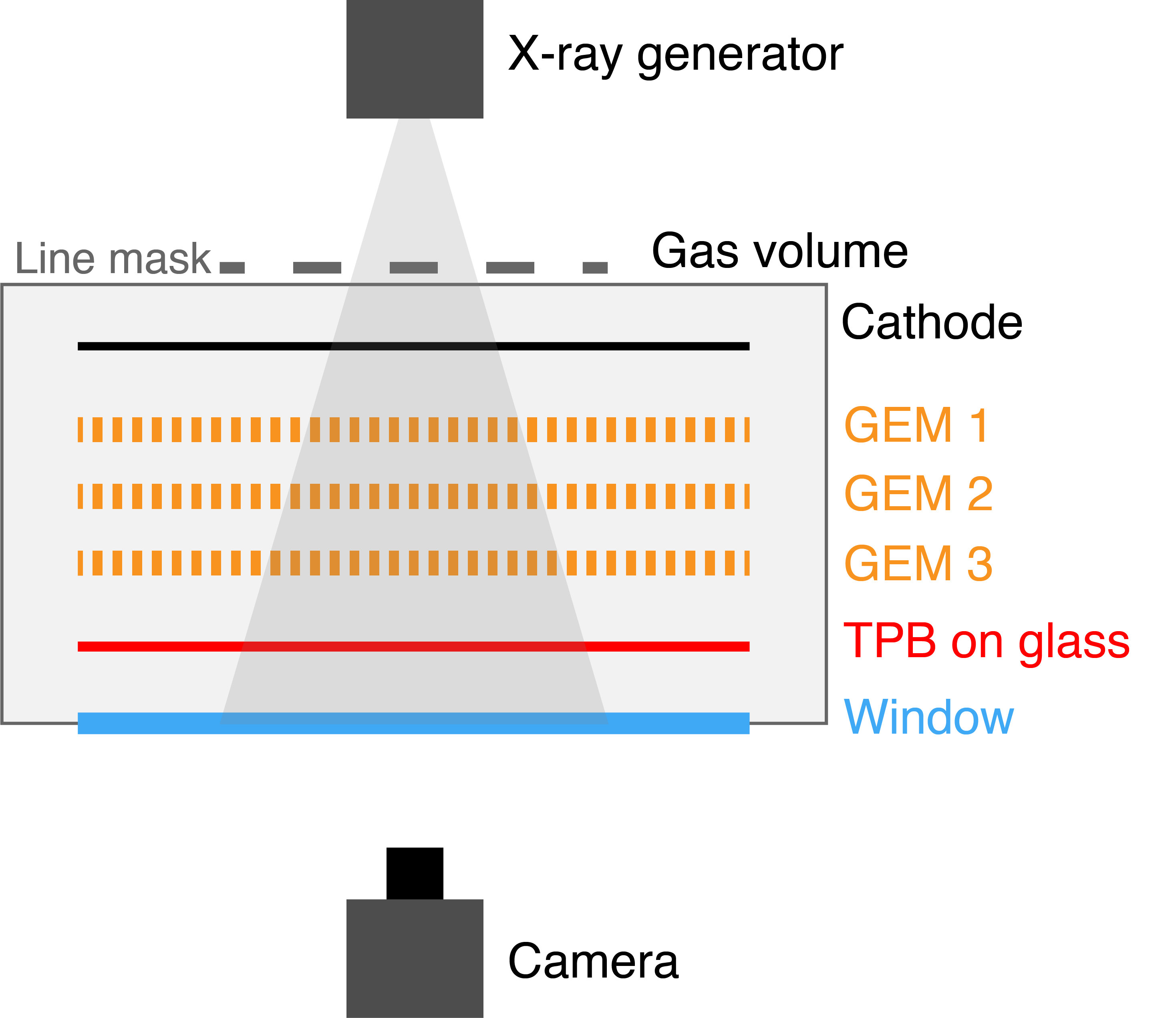}
        \caption{Triple-GEM setup}
    \end{subfigure}%
    ~ 
    \begin{subfigure}[t]{0.5\textwidth}
        \centering
        \includegraphics[height=2in]{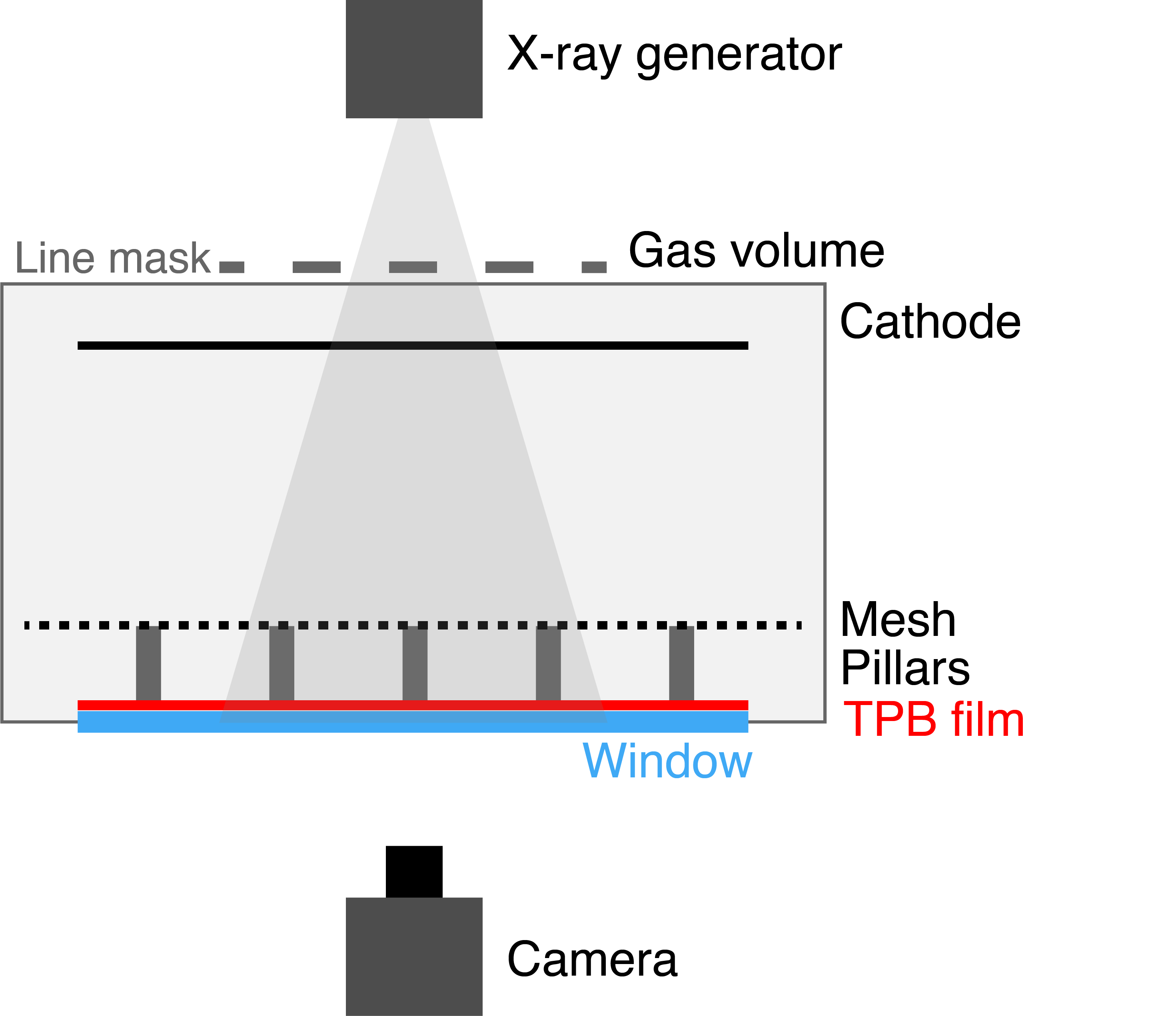}
        \caption{Micromegas setup}
        \label{fig:wlsSetupMM}
    \end{subfigure}
    \caption{Optically read out detectors with semi-transparent TPB WLS. Detector was irradiated with an X-ray tube. A camera placed after the amplification stage and WLS plate was used to read out images of a line mask placed in front of the detector. Schematics are not to scale. The distance between the GEMs was set to 2\,mm. The drift region thickness was also set to 2\,mm.}
        \label{fig:setups}

\end{figure*}

A triple-GEM detector enclosed in a gas vessel with a transparent window was employed to characterise spatial resolution in an X-ray radiography setup, as illustrated in Fig.~\ref{fig:wlsSetupMM}.
 The detector was irradiated with an X-ray generator placed at a distance of approximately 1\,m from the detector. A resistive voltage divider was used to power the GEM stack with equal potential differences across GEMs 1 and 2 as well as in the transfer gap between the GEMs. A Cu-on-Kapton foil cathode was positioned at a distance of 2\,mm above the first GEM. A wavelength-shifting plate, consisting of a TPB film deposited on a glass substrate, was positioned at various distances below the GEM stack. The TPB layer was produced by thermal evaporation and had a thickness of approximately 860\,nm, resulting in partial transparency to visible light, with a transmission of about 50\% at 800\,nm.

A bulk glass Micromegas \cite{glassMicromegas} detector was used to demonstrate the possibility to integrate the wavelength shifter directly on the anode plane. A glass plate coated with an ITO film with a nominal surface resistivity of 4\,$\Omega / \square$ was used as substrate for the Micromegas. Pillars with a pitch of 6\,mm were structured in 128\,$\mu$m thick coverlay by photolithpography.
TPB was deposited on the ITO-coated anode plane thus covering the full anode surface as well as the coverlay pillars. Due to the apparent high resistivity of TPB, no significant leakage currents were observed between the grounded mesh and the ITO anode. Despite this, the Micromegas operation was not affected and high electric fields could be applied.
A stainless steel micromesh (woven, calendared 18\,$\mu$m wires, 45\,$\mu$m holes) was stretched on a frame and pressed against the pillars. The mesh was grounded, while the ITO anode was biased to positive high voltage. A cathode foil (Cu on Kapton) was placed at a distance of 2\,mm above the micromesh.

To compare the effect of wavelength shifting on the achievable spatial resolution, X-ray radiography images of a line pair phantom were recorded as shown in figure \ref{fig:setups}. The detector was operated with an Ar/CF$_{4}$ (80/20\%) gas mixture at ambient pressure which was flushed at 5 litres per hour. The emission spectrum of this gas mixture features strong molecular emission bands from CF$_{4}$ in the UV and VIS ranges \cite{cf4GEMScintillation}. 

X-ray radiography images of the line pair mask were recorded with 30\,s exposure time. An X-ray generator with a Cu target was used at a setting of 20\,kV and varying currents from 0.1\,mA to 40\,mA. Background images were acquired with the same settings and subtracted to account for the baseline of the image sensor and any potential ambient light leakage. A single 30\,s exposure was used for each data point. White images without the line pair mask were recorded and background subtracted images were divided by white images to correct for non-uniform irradiation caused by the X-ray beam profile and any potential non-uniformities of the response of the detector.

\begin{figure*}[t!]
    \centering
    \begin{subfigure}[t]{0.5\textwidth}
        \centering
        \includegraphics[height=2in]{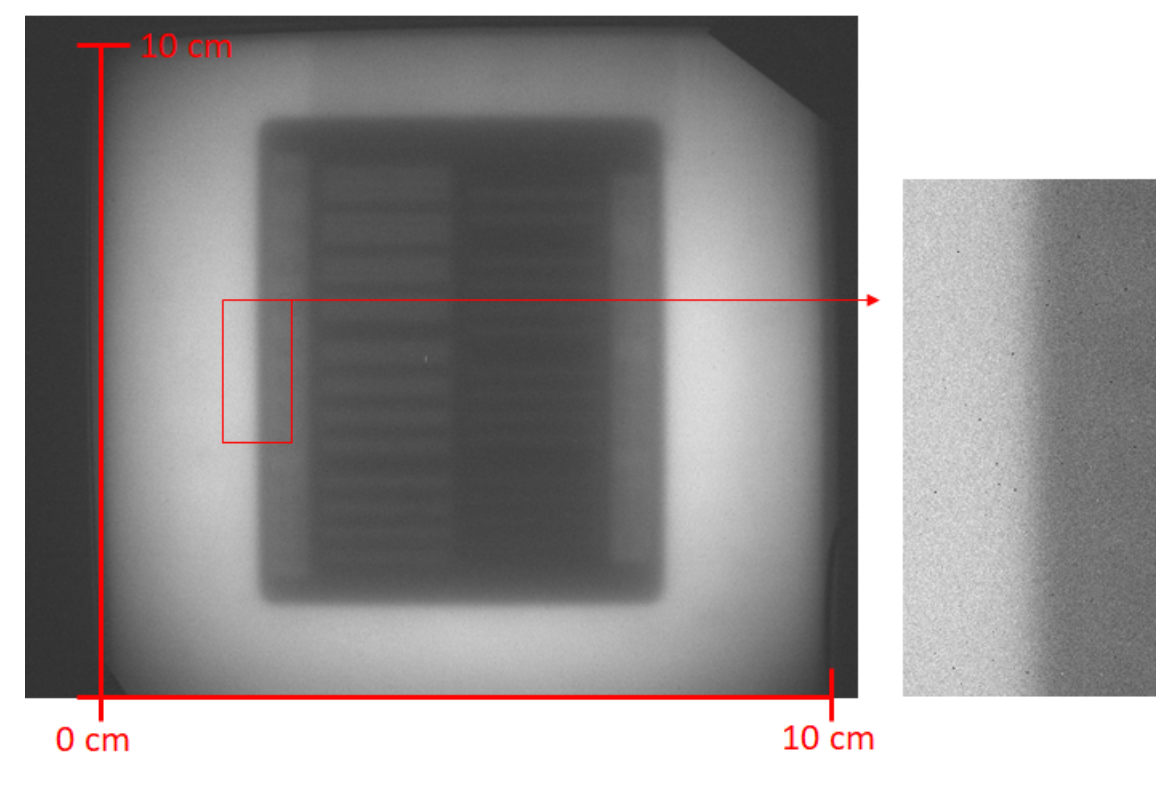}
        \caption{Extraction of transition from bright to dark region from images.}
        \label{fig:projectionAnalysis}
    \end{subfigure}%
    ~ 
    \begin{subfigure}[t]{0.5\textwidth}
        \centering
        \includegraphics[height=2in]{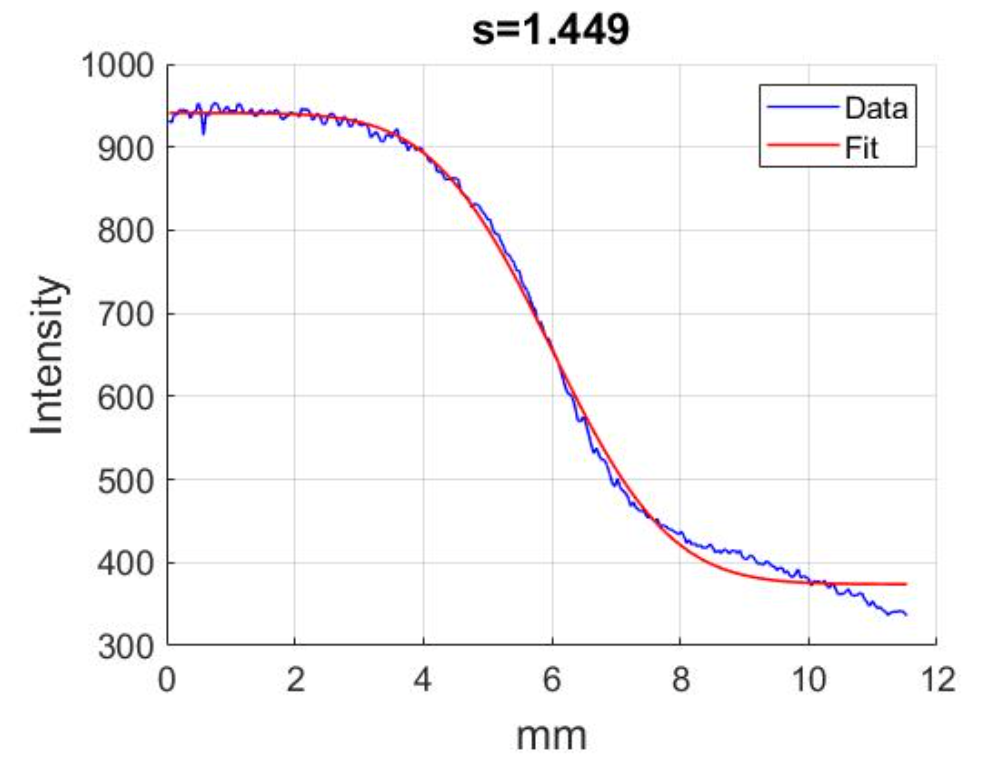}
        \caption{Fit of projected intensity with error function.}
        \label{fig:stepFit}
    \end{subfigure}
    \caption{Procedure to determine achievable resolution from transition from bright to dark region. Images taken from \cite{NummiSummerStudentReport}.}
        \label{fig:setups}

\end{figure*}

The achievable resolution was determined by the sharpness of the transition from a bright to a dark region. For this, a transition region was extracted from an image as shown in figure \ref{fig:projectionAnalysis}. The intensity from the extracted image region was projected onto one dimension to obtain a line profile of the transition. The line profile was fitted with a step function defined by the error function as shown in figure \ref{fig:stepFit}:
\begin{equation}
y(x) = y_0 + \frac{A}{2} \,\mathrm{erf}\!\left( \frac{x - x_0}{\sqrt{2}\,\sigma} \right),
\end{equation}
where $\mathrm{erf}$ is the error function $\mathrm{erf}(x) = \frac{2}{\sqrt{\pi}} \int_0^x e^{-t^2} \, dt$ Free parameters are $y_0$, $A$, $x_0$, and $\sigma$. The parameter $\sigma$ was used as width of the step function and thus as indication of the achievable resolution.

Scintillation light spectra, used to confirm wavelength shifting, were recorded using an Ocean Optics FLAME S UV–VIS spectrometer. Light was collected with a 5\,mm-diameter collimating lens and coupled via an optical fiber to the 200\,$\mu$m entrance slit of the spectrometer. Spectra were acquired with an exposure time of 30\,s and background spectra acquired without irradiation of the detector were subtracted. A calibration of the response of the spectrometer was performed with a calibration lamp and applied to acquired spectra

\begin{figure}[h!]
\centering
\makebox[\linewidth]{\includegraphics[width=0.7\textwidth]{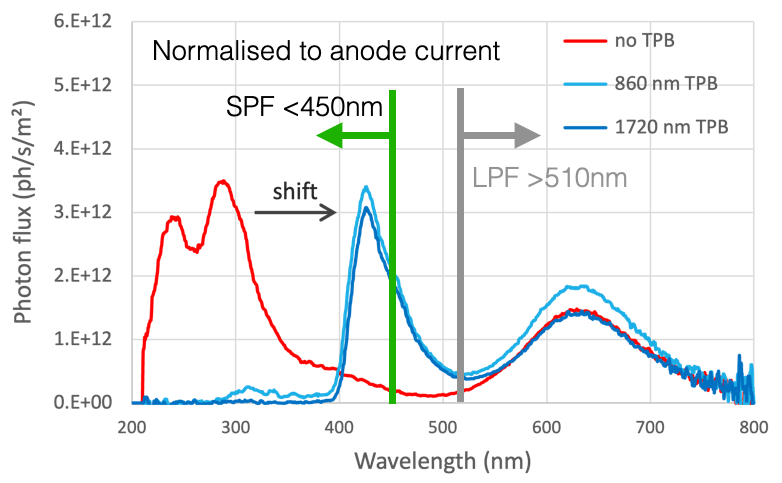}}
\caption{CF$_{4}$ emission and reemission spectrum from TPB. The cutoff wavelengths of the used shortpass filter (SPF) and longpass filter (LPF) are indicated. Scintillation spectra were normalised to the anode current.}
\label{WLSShiftingSpectrum}
\end{figure}

In proportional mode, the addition of a few percent of a molecular quencher (CF$_{4}$) is sufficient to make the scintillation spectrum dominated by the quencher’s molecular light \cite{cf4Scintillation} and CF$_{4}$ emits in both UV and VIS bands. The UV band is efficiently absorbed by the TPB and subsequently re-emitted at the higher wavelength range around 450\,nm. The visible scintillation band of CF$_{4}$ centered around 630\,nm is not shifted by TPB and is partially transmitted by the semi-transparent WLS plate and directly recorded by the camera.
Thus, the light recorded by the camera is composed of two components: one re-emitted by TPB and one directly emitted by CF$_{4}$. To differentiate the two components short and long pass filters were placed in front of the lens of the camera. A shortpass filter with a cutoff wavelength of 450\,nm was used to select only the TPB emitted light. The direct UV light from CF$_{4}$ was suppressed by the glass plate with a cutoff wavelength around 350\,nm. The emission specturm and the cutoff values of the filters used for the measurements is shown in figure \ref{WLSShiftingSpectrum}. The spectrum shown is normalised to the anode current, i.e. the total amount of charge carriers collected.

A CCD camera with a resolution of 6 megapixels (Retiga R6) was used with an f/0.95 lens with a focal length of 25\,mm and an addition +3 dioptre lens to enable a focusing distance of approximately 30\,cm.

\section{Solid wavelength shifter readout}
\label{sec:spectra}

A line-pair mask, consisting of a 50\,$\mu$m-thick Pb foil with lines of varying widths, was placed in front of the device, as shown in Fig.~\ref{fig:setups}. The Pb mask was encapsulated in plastic. The detector was illuminated with X-rays from a generator operated at 20\,kV energy. Background images recorded without X-rays were subtracted from acquired images to account for the baseline of the imaging sensor and any potential ambient light leakage. The shortpass filter was used to select the light emitted by the TPB WLS.

\subsection{Triple-GEM detector with TPB plate}
\label{sec:spectra}

\begin{figure*}[t!]
    \centering
    \begin{subfigure}[t]{0.5\textwidth}
        \centering
        \includegraphics[height=2in]{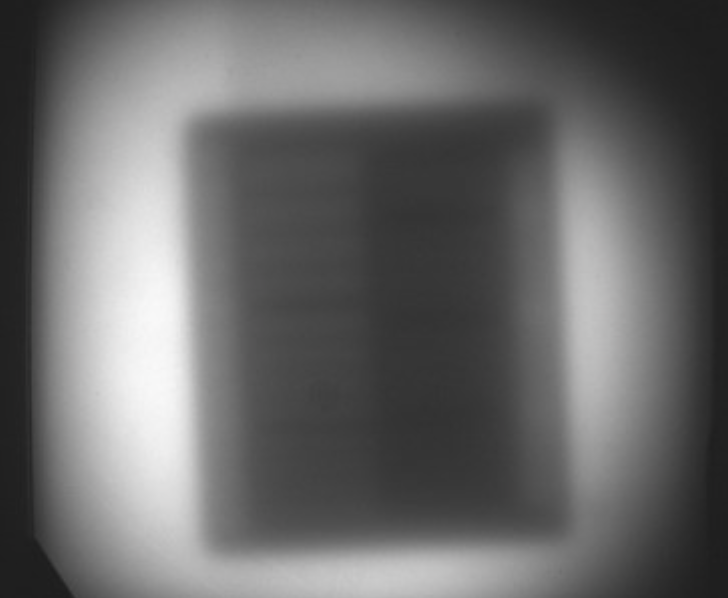}
        \caption{2\,mm gap between GEM and TPB}
    \end{subfigure}%
    ~ 
    \begin{subfigure}[t]{0.5\textwidth}
        \centering
        \includegraphics[height=2in]{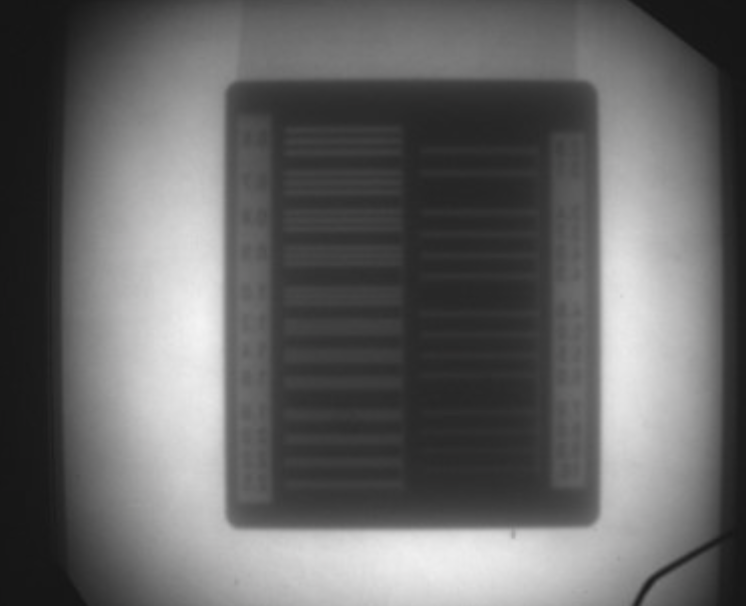}
        \caption{No gap between GEM and TPB}
    \end{subfigure}
    \caption{Effect of WLS location on image blurring recorded with TPB plate placed below triple GEM. The shortpass filter was used for the shown images.}
    \label{fig:GEMWLSlocation}
\end{figure*}

The effect of the distance between the site of light emission and the wavelength shifting film was studied by placing a TPB-coated glass plate at varying distances below a triple-GEM stack. Distances of 2\,mm, 1\,mm, 0.5\,mm and 0\,mm between the last GEM and the TPB plate were investigated. The camera was focused on the back of the TPB-coated glass plate by attaching a tape with a cross on the back of the glass which was removed after focusing. A comparison between the blurring of 2\,mm gap and no gap between last GEM and TPB plate is shown in figure \ref{fig:GEMWLSlocation}.

The distance between the original light production at the GEM and the absorption and re-emission at the TPB plate clearly results in a strong blurring which degrades spatial resolution. Scintillation light is emitted isotropically during avalanche multiplication. The fraction of the light which reaches the TPB layer is absorbed and reemitted isotropically at a longer wavelength. Thus, larger distances between between the amplification structure and the WLS result in increased blurring.
A comparison of the edge-spread-function (ESF) achieved in different configurations is reported in table \ref{tableComparison}.

\begin{table}
\begin{center}
\begin{tabular}{ |c|c|c| } 
 \hline
  & Gap to WLS (mm) & ESF (mm) \\ 
  \hline
 \multirow{3}{4em}{Triple GEM}  & 2 & 2.09 \\ 
  & 1 & 1.43 \\ 
  & 0.5 & 0.8 \\ 
  & 0 & 0.46 \\ 
 \hline
 Micromegas & 0 & 0.22 \\ 
 \hline
\end{tabular}
\caption{Recorded edge-spread-function (ESF) for triple GEM and MM with WLSs.}
\label{tableComparison}
\end{center}
\end{table}

At a separation of 2\,mm between the last GEM and the WLS plate, the measured edge spread function (ESF) exhibited a width exceeding 2\,mm. In contrast, in the optimal configuration with no additional separation between the bottom electrode and the WLS plane, a spatial resolution of 0.46,mm was achieved. 
While the expected result that the GEM in well configuration achieves the best spatial resolution demonstrates the possibility to achieve good imaging resolution with a GEM-based amplification structure combined with a solid WLS, glass Micromegas might offer even better resolution due to their single stage amplification, finer primary charge sampling due to the pitch of the mesh wires compared to the larger hole pitch of the GEM and the possibility to coat the WLS layer directly on the anode.

\subsection{Micromegas with TPB film}
\label{sec:mmTPB}

A bulk Micromegas integrated on a glass substrate and Indium-Tin-Oxide (ITO) layer as anode was used to evaluate the achievable spatial resolution with a WLS directly coated on the anode. The Micromegas featured an active area of 8\,cm x 8\,cm. Two methods for fabricating glass Micromegas with an embedded WLS were explored: As a first method, TPB was deposited on ITO-coated glass plates before the bulking process \cite{bulkMM}. In this case, however, the chemical baths required for cleaning and etching the substrate during the bulking process led to the damage of the TPB layer. Therefore, a second method was adapted, in which a 860\,nm thick film of TPB was deposited on the already bulked Micromegas by thermal evaporation. This resulted in not only the anode but also the mesh and some coverlay regions being coated with TPB, which did not present an issue since TPB is non conductive. Electrical tests after WLS coating confirmed that cathode, mesh and anode were still electrically insulated. In the presented tests, no significant charging up effects were observed although this may be a concern for prolonged operation. To mitigate possible charging up in high rate conditions, a semi-transparent thin coating could be applied on top of the WLS such as Ti or Cr layers of a few nm thickness.

In the presented studies, the TPB layer was deposited on the anode side to allow UV light to be converted to the visible range before passing through the glass substrate. Alternatively, the use of a UV-transparent anode and substrate such as MgF$_2$ would allow for a deposition of the WLS on the opposite side of the substrate thus eliminating concerns about charging up and damage due to discharges.

\begin{figure}[h!]
\centering
\makebox[\linewidth]{\includegraphics[width=0.5\textwidth]{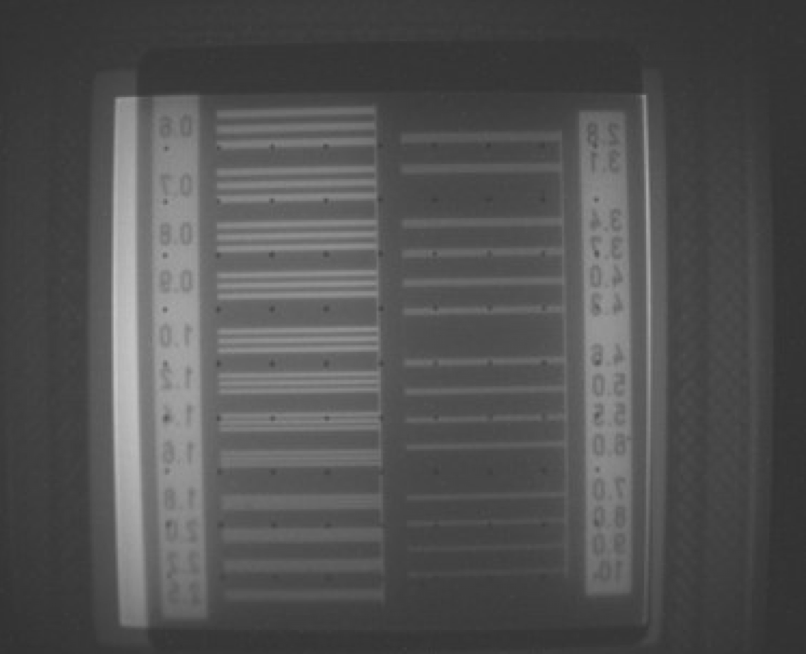}}
\caption{Line pair mask image recorded with Micromegas with TPB-coated anode and shortpass filter.}
\label{mmTPBImage}
\end{figure}

An image of the line pair phantom acquired with the TPB-coated glass Micromegas is shown in figure \ref{mmTPBImage}. Significant improvement over the images acquired with a triple-GEM amplification stage can be seen. The pillars of the Micromegas show up as a regular pattern of ineffective (black) areas. When fitting an error function to the transition between bright and dark regions, an ESF of 0.22\,mm was determined which is about a factor of two better than the GEM+WLS configuration with no gap. While the triple GEM stack allows for larger amplification factors compared to the single stage multiplication of the glass Micromegas, it also adds additional diffusion through the two transfer gaps which limits the achievable resolution.

\section{Alternative optical readout gas mixtures}
\label{sec:otherGases}

\begin{figure}[h!]
\centering
\makebox[\linewidth]{\includegraphics[width=0.9\textwidth]{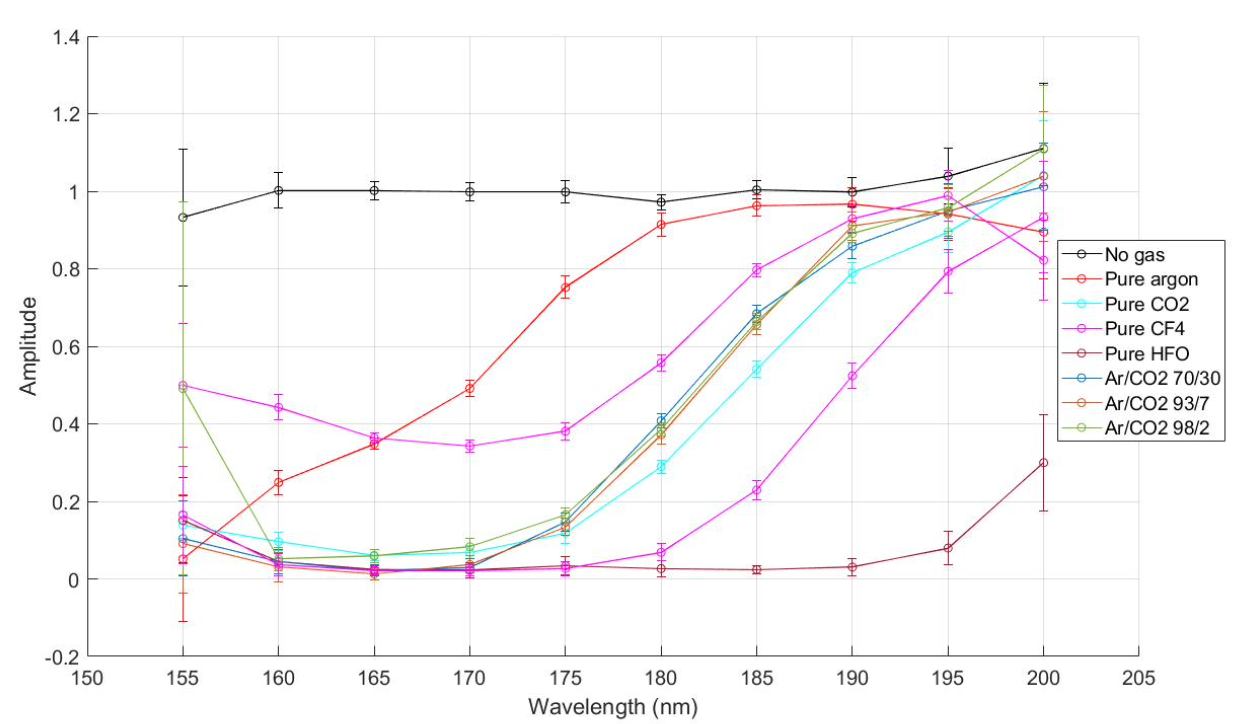}}
\caption{Transparency of different gases in conditions commonly used in gaseous detectors.}
\label{TransparencyComparison}
\end{figure}

The possibility to perform optical readout of MPGDs with gas mixtures other than CF$_{4}$-based mixtures is attractive to increase the versatility of this readout approach while simultaneously minimising the dependence on CF$_{4}$ which is a strong greenhouse gas under increasing restrictions. In addition to featuring UV scintillation light emission which may be converted to visible light by WLS layers, candidate gases should also feature sufficient transparency to the emitted light spectrum to allow light to reach the WLS layer. The VUV transparency of some commonly used detector gases and mixtures was measured between 155 and 200\,nm as shown in figure \ref{TransparencyComparison}. Measurements were conducted with the ASSET setup which consists of a deuterium lamp, a VUV monochromator and focusing optics \cite{lisowska2024photocathodecharacterisationrobustpicosec}. The light is split into two paths, a reference path for monitoring the stability of the light source and a measurement path. In both paths, the light intensity is measured by calibrated CsI PMTs. The different gases were filled into the measurement chamber and pressure was monitored by a capacitive gauge with a precision of 1\,mbar. The measurements were conducted in sealed mode. Due to the lower light intensity at the short and long limits of the shown range, the error bars of the transparency measurements increase in these regions.

While pure Ar features a higher transparency, the VUV transmission of Ar/CO$_{2}$ mixture appears to be dominated by the presence of CO$_{2}$. 

\begin{figure}[h!]
\centering
\makebox[\linewidth]{\includegraphics[width=0.6\textwidth]{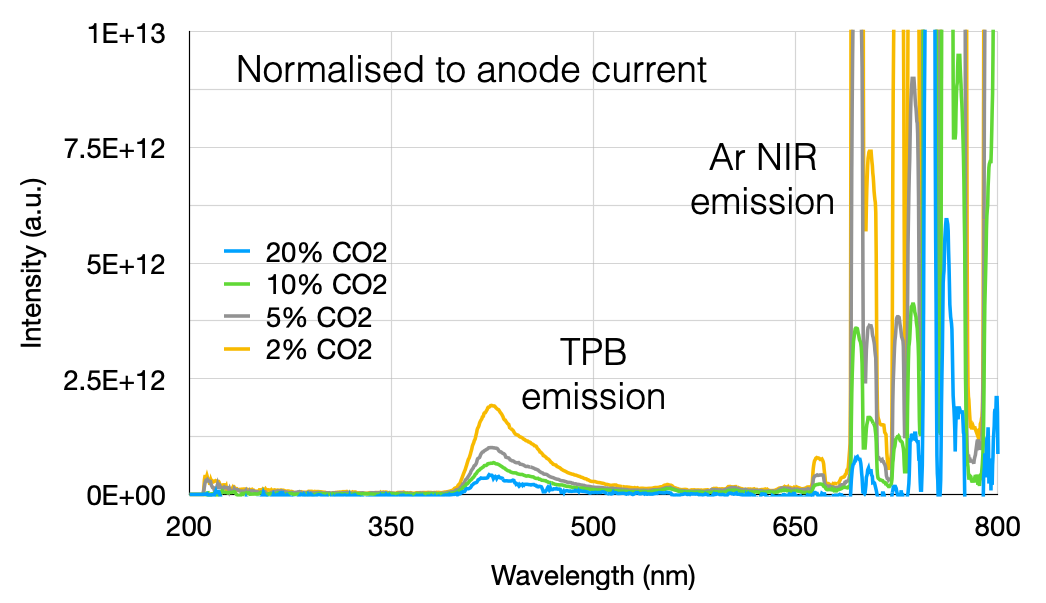}}
\caption{Scintillation light spectra of Ar/CO$_{2}$ mixtures with different mixing ratios.}
\label{ArCO2-Emission}
\end{figure}

In addition to the mixing ratio between Ar and CO$_{2}$, also the distance light has to travel in that mixture before reaching the WLS layer impacts the resulting transmission. The possibility to convert UV light from Ar/CO$_{2}$ mixtures to visible light with a TPB layer was evaluated in a GEM-based detector with a distance of 2\,mm between the bottom electrode of the GEM and the WLS layer. A comparison of different gas mixtures ranging from 2-20\% CO$_{2}$ content is shown in figure \ref{ArCO2-Emission}. The data is normalised to the anode current, i.e. the total amount of charge carriers produced during avalanche amplification, to compare the absolute intensity of the resulting visible light component. The anode current was measured by averaging 20 current readings and subtracting the baseline current offset obtained without the radiation source. The acquired spectra were then divided by the obtained anode current measurement. In addition to the visible light component around 450\,nm emitted by TPB, near-infrared (NIR) lines of Argon are visible, the intensity of which is correlated to the fraction of Ar in the mixtures. The intensity of the visible light emitted by TPB increases with lower CO$_{2}$ content, with the intensity obtained with 2\% CO$_{2}$ about five times higher than the one obtained with 20\% CO$_{2}$. This shows that low CO$_{2}$ content is advantageous to obtain ample scintillation light emission. However, higher quenching gas content may enable larger maximum charge gain and could thus help to obtain a higher absolute amount of scintillation light. In the present study, the detector was operated at comparable gains and maximum achievable gain was not evaluated.

\section{Micromegas with TPB and isobutane based gas mixture}
\label{sec:IsoMicromegas}

\begin{figure}[htpb]
    \centering
    \begin{minipage}[c]{0.39\textwidth}
        \includegraphics[width=\textwidth]{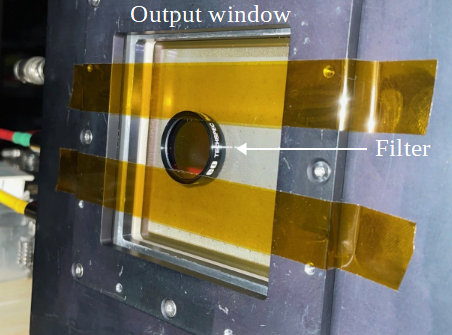}
    \end{minipage}
    \begin{minipage}[c]{0.6\textwidth}
        \includegraphics[width=\textwidth]{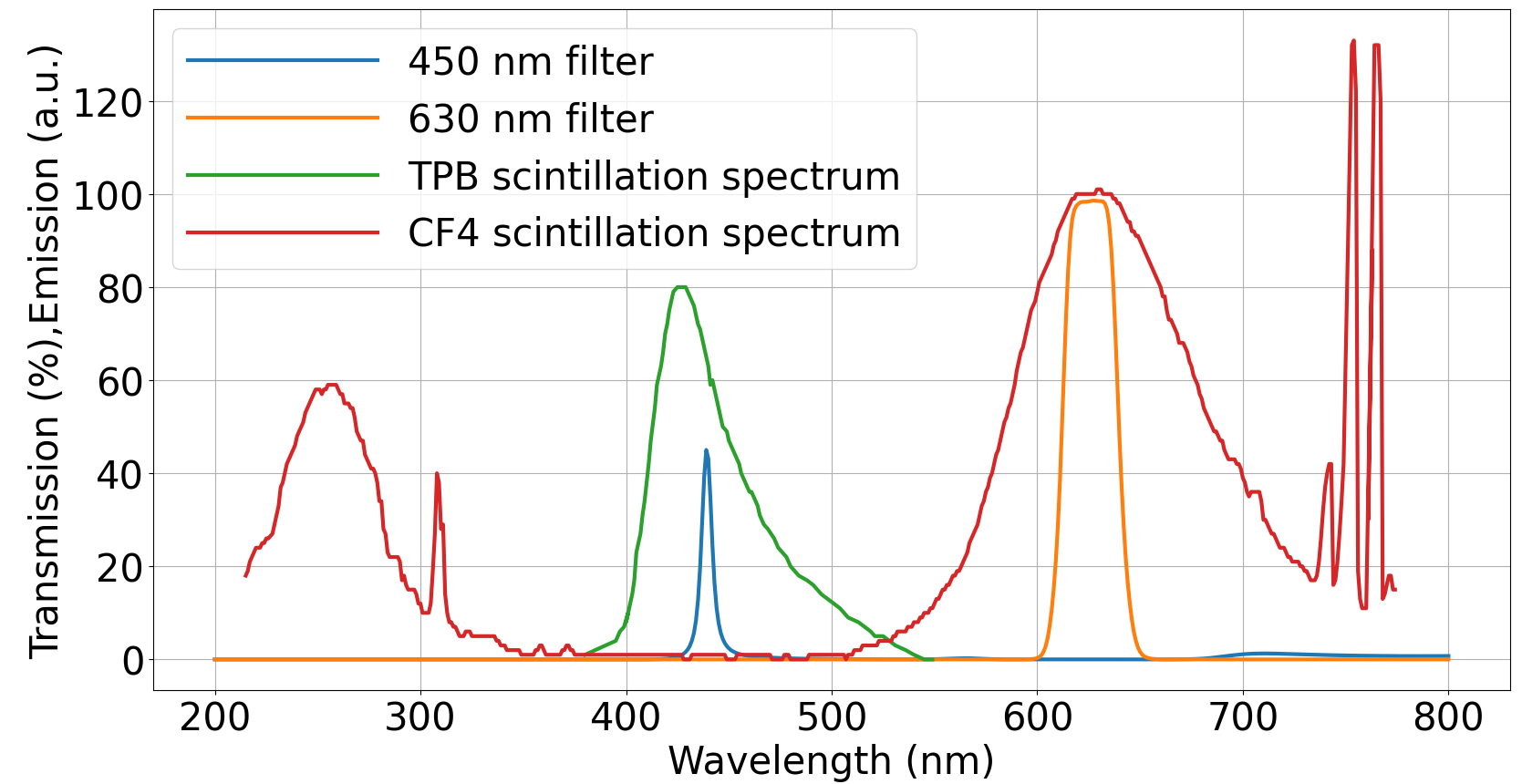}
    \end{minipage}
    \caption{Picture of the glass Micromegas detector front window with an optical filter (left). Scintillation spectrum of an Argon/CF$_4$ based gas mixture (red plot) and TPB (green plot). 450\,nm filter (blue plot) and 630\,nm filter (orange plot) wavelength dependent transmissions.}

\label{Fig:TPBTest}
\end{figure}

A test has been performed to analyze the light produced by the TPB based glass Micromegas detector. Two different optical filters have been placed at the output window of the detector (Figure\,\ref{Fig:TPBTest}, left). As shown on (Figure\,\ref{Fig:TPBTest}, right), one filter is only transparent to the light scintillated by a CF$_4$ based gas mixture (red plot), with a transmission peak at 630\,nm (orange plot). The other filter is only transparent to the light scintillated by the TPB (green plot), with a transmission peak at 450\,nm (blue plot). This way, if light crosses the 630\,nm filter, CF$_4$ is involved, and if it crosses the 450\,nm filter, TPB is involved.

\begin{figure}[htpb]
    \centering
    \includegraphics[width=0.488\textwidth]{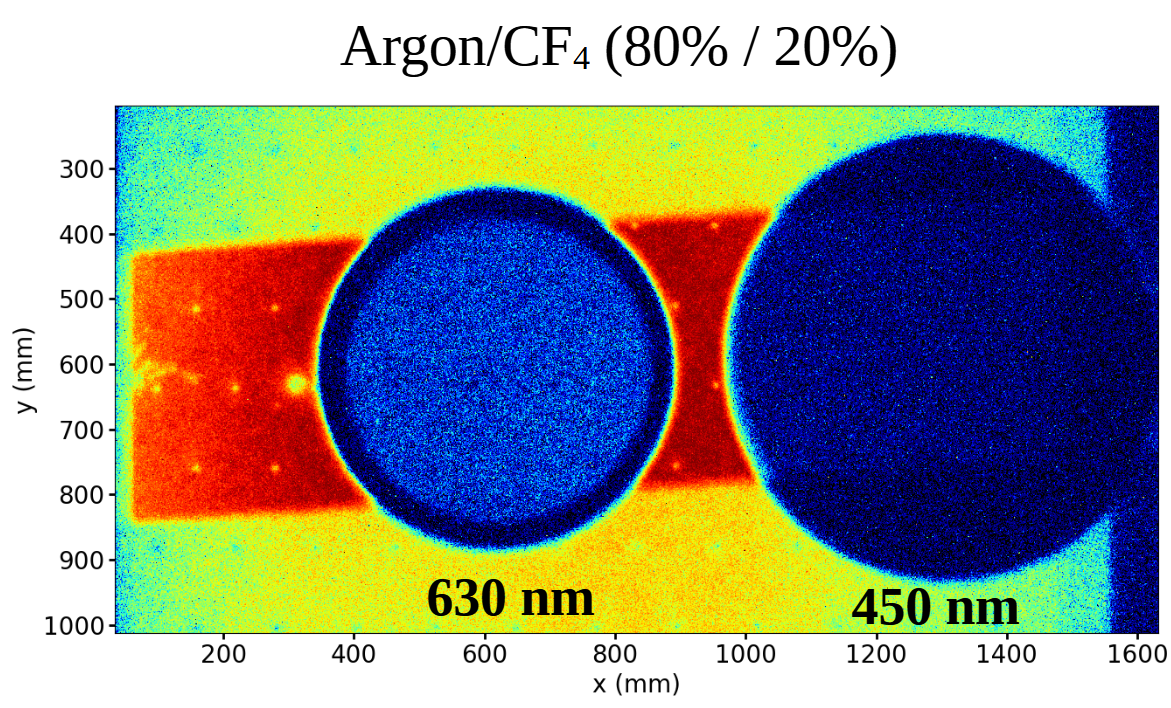}
    \includegraphics[width=0.49\textwidth]{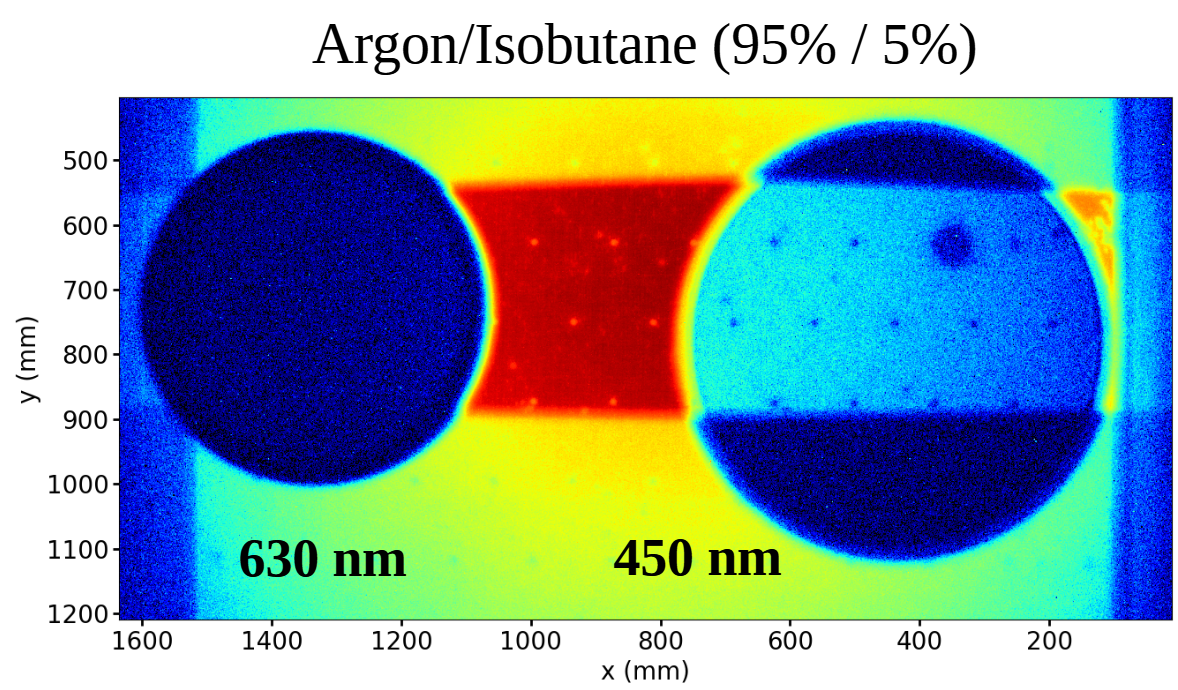}
    \caption{Light intensity frames recorded with the qCMOS camera. The TPB based glass Micromegas is operated with an Argon/CF$_4$ (80\%/20\%) gas mixture (left) and with an Argon/Isobutane (95\%/5\%) gas mixture (right). The 630\,nm filter and the 450\,nm filter are identified. Images were acquired with a Hamamatsu Orca Quest camera \cite{orcaQuest}.}

\label{Fig:WLSFilters}
\end{figure}

The TPB based glass Micromegas detector has been tested with a standard mesh and a 128\,µm amplification gap thickness, using a $^{55}$Fe source. First, the detector was operated with an Argon/CF$_4$ (80\%/20\%) gas mixture at a amplification field of 49\,kV/cm. As shown in Figure\,\ref{Fig:WLSFilters} (left), only the 630\,nm filter lets the light pass, indicating that the scintillation light indeed comes from the CF$_4$ based gas mixture. Then, the detector was operated with an Argon/Isobutane (95\%/5\%) gas mixture at a amplification field of 37\,kV/cm. Inversely, only the 450\,nm filter lets the light pass in Figure\,\ref{Fig:WLSFilters}, showing that the light comes from the TPB scintillation. 

These results demonstrate that visible scintillation can be achieved in the glass Micromegas detector using TPB in combination with an Ar/Isobutane (95/5\%) gas mixture.

\section{Conclusions}
\label{sec:conclusions}
Optical readout of MPGDs with solid WLS layers was shown to achieve high spatial resolution with the best value of ESF of 0.22\,mm achieved by bulk Micromegas with a TPB layer coated onto the anode. The minimisation of the distance between the site of scintillation light emission and the wavelength shifting layer decreases image blurring and improves spatial resolution. For a triple-GEM with the WLS directly placed at the bottom of the last GEM, a ESF of 0.46\,mm could be obtained. Ar/CO$_{2}$ gas mixture may be used in combination with TPB as solid wavelength shifter with lower CO$_{2}$ contents increasing visible scintillation light output.
The presented study demonstrates the possibility to use solid wavelength shifters such as TPB with other gas mixtures than the typical CF$_{4}$-based mixtures associated with optical readout. Additional studies of alternative gas mixtures are planned to inform choices for optically read out detectors which do not rely on CF$_{4}$ and increase the versatility of this readout approach. 
As a robust alternative to TPB, PEN film may be used as solid wavelength shifter although it achieves lower wavelength shifting efficiency. PEN sheets instead of thin TPB layers may feature a higher robustness against prolonged operation in MPGDs including electron bombardment and discharges. The integration of PEN as wavelength shifting substrate in bulk Micromegas is currently being developed and the achievable efficiency will be compared to TPB-coated glass Micromegas.


\bibliographystyle{JHEP}
 \bibliography{biblio.bib}

\end{document}